# Finite-difference frequency-domain method for the extraction of effective parameters of metamaterials

João T. Costa, Mário G. Silveirinha[*], Stanislav I. Maslovski

University of Coimbra, Department of Electrical Engineering – Instituto de Telecomunicações, Portugal, joao.costa@co.it.pt, mario.silveirinha@co.it.pt, stas@co.it.pt

**Abstract**

Here, we report a numerical implementation of the nonlocal homogenization approach recently proposed in [M. G. Silveirinha, Phys. Rev. B **75**, 115104 (2007)], using the finite-difference frequency-domain method to discretize the Maxwell-Equations. We apply the developed formalism to characterize the nonlocal dielectric function of several structured materials formed by dielectric and metallic particles and in particular, we extract the local permittivity, permeability and magnetoelectric coupling parameters when these are meaningful. It is shown that the finite differences frequency domain implementation of the homogenization method is stable and robust, yielding very accurate results.

---

[*] To whom correspondence should be addressed: E-mail: mario.silveirinha@co.it.pt

# I. INTRODUCTION

The interest in structured materials (metamaterials) with atypical electromagnetic properties has received significant attention in recent years. Such materials, may interact with electromagnetic waves in a controlled and desired way, and provide unique electromagnetic responses, distinctively different from those of conventional dielectrics and metals. For example, it has been shown that by tailoring the microstructure of conventional materials, so that the geometry of the inclusions and their spatial arrangement is thoughtfully chosen, it may be possible to synthesize novel media with unconventional properties, such as simultaneously negative permittivity and permeability [1], extreme anisotropy [2], permittivity near zero [3, 4], or broadband anomalous dispersion with no loss [5]. These novel materials may permit interesting applications and effects such as imaging not limited by diffraction [6], boost the sensitivity of magnetic resonance imaging [7], negative refraction [8], noninvasive sensing [9], or the realization of ultra-subwavelength waveguides and resonators [10, 11, 12].

Similar to conventional materials, due to the relatively small dimensions of the inclusions as compared to the wavelength of radiation, the electromagnetic wave propagation in metamaterials can be conveniently described using homogenization techniques. Indeed, in general it is difficult or impractical to model all the minute microscopic details of the basic inclusions that form a specific metamaterial, since this may require extensive computational efforts, sometimes beyond the available computational resources. Homogenization methods greatly simplify propagation and radiation problems, by considering that the structured material may be regarded as a continuous medium described by only a few *effective parameters*, typically the permittivity and permeability.

Due to this reason in recent years significant efforts have been devoted to the extraction of the effective parameters of metamaterials. The homogenization of heterogeneous structures is a topic with a long history [13], with key contributions from researchers like Lorentz, Planck, and Oseen, and more recently Milton, Bergman, among others. The simplest homogenization techniques are based on the use of mixing-formulas such as the classic Clausius-Mossotti (CM) formula. The CM-formula can be quite accurate in some scenarios, but is limited by the fact that it considers that the volume fraction of the inclusions is small, which in practice is rarely the case. Valuable information about the effective response of a periodic material can also be obtained from its electromagnetic band structure, but such procedure is of limited applicability in case of losses and in electromagnetic band-gaps [14]. An alternative averaging approach based on field summations inside the material was proposed in Ref. [15]. The computation of the effective parameters of metamaterials based on the inversion of computed or measured scattering data [16, 17] is indisputably the most popular method to characterize composite media nowadays. However, such approach has several problems, namely the extraction method may yield multiple solutions, making it cumbersome to determine the correct branch and, in some cases the extracted parameters may be unphysical [18 , 19].

Recently, we have introduced a systematic and completely general homogenization method to extract the effective parameters of composite media [20, 21]. The method is based on the solution of a source-driven numerical problem, and thus does not involve the computation of eigenmodes. The solution of the homogenization problem formulated in Ref. [20] yields unambiguously the effective parameters of the composite medium. In contrast with previous works, which typically assume that the electrodynamics of the material can be described in terms of bianisotropic constitutive relations, the method proposed in Ref. [20] describes the metamaterial in terms of a nonlocal dielectric function, i.e. using a spatially dispersive model. A material with spatial dispersion is characterized by the fact that the polarization acquired by the inclusions does not depend exclusively on the macroscopic (i.e. averaged) electric field in the immediate vicinity of the particle, but depends also on the macroscopic electric field at distances larger than the characteristic dimension of the basic cell. Consequently, in the space domain, the electric displacement vector **D** and the electric field **E** are typically related through a spatial convolution, and thus for plane waves (spectral domain) the dielectric function becomes a function of both the frequency of operation and of the wave vector.

An important property of the homogenization approach of Ref. [20] is that the constitutive relations implicit in the spatially dispersive model are very general (more specifically they can model an arbitrary composite material such that the electric displacement vector is related to the macroscopic electric field through a spatial convolution), and thus include as a particular case the conventional constitutive relations that characterize the material using effective permittivity and permeability tensors, and possibly magneto-electric tensors (bianisotropic model). In particular, as demonstrated in Ref. [20], if meaningful, the conventional local parameters (permittivity and permeability) may be extracted from the nonlocal dielectric function by computing the derivatives of the dielectric function with respect to the wave vector. Moreover, nonlocal homogenization methods assume particular importance in the characterization of metamaterials with strong spatial dispersion [5, 12, 22], which cannot be homogenized with classical methods. It has been shown recently that such materials may have quite interesting applications such as the transport and manipulation of the electromagnetic fields in the nanoscale [2 , 23], or the anomalous dispersion of light colors [5].

In Ref [20] it was shown that the dielectric function of a metamaterial may be numerically computed using the Method of Moments (MoM). Even though such procedure is completely general, it is well known that the MoM is mainly suitable for the characterization of metallic structures, being less efficient in the analysis of dielectric structures, where finite difference methods are generally much more versatile and powerful [24]. The aim of this paper is to demonstrate that the homogenization problem can be efficiently solved using a finite difference frequency domain (FDFD) formalism [25].

This paper is organized as follows. In Sec. II we review the homogenization approach introduced in Ref. [20] and explain how the homogenization problem can be solved by discretizing the Maxwell Equations using the FDFD method. In Sec. III we validate the proposed formalism, and illustrate its application to several metamaterials formed by dielectric inclusions. The extracted local effective parameters ($\varepsilon_{eff}$ and $\mu_{eff}$) are compared with those obtained with other homogenization methods, and in particular we discuss with details the physics of plasmonic metamaterials that exhibit artificial magnetism. Finally, in Sec IV, the conclusions are drawn.

In this work, it is assumed that the fields are monochromatic with a time dependence of the form $e^{j\omega t}$.

## II. NONLOCAL HOMOGENIZATION USING THE FDFD METHOD

Here, we describe an FDFD [25] solution of the homogenization problem formulated in Ref. [20]. To begin with, we present an overview of the homogenization method, explaining its principles and how it can be used to extract the effective parameters (e.g.

$\varepsilon_{eff}$ and $\mu_{eff}$) of metamaterials, and subsequently we apply the FDFD formalism to solve the pertinent source-driven problem.

For simplicity, without loss of generality, it is assumed that the structured material under study is a two-dimensional metal-dielectric crystal, obtained by translating a two-dimensional unit cell $\Omega$ along the primitive vectors $\mathbf{a}_1$ and $\mathbf{a}_2$. The periodic material can have dielectric and/or metallic inclusions, with the magnetic permeability equal to $\mu_0$ and relative permittivity $\varepsilon_r = \varepsilon_r(\mathbf{r},\omega)$.

### A. Overview of the homogenization method

The method presented in Ref. [20] permits the extraction of the effective parameters of a generic periodic composite material formed by nonmagnetic or metallic inclusions. The aim of the method is to calculate the nonlocal dielectric function $\overline{\overline{\varepsilon}}_{eff} = \overline{\overline{\varepsilon}}_{eff}(\omega,\mathbf{k})$ of the metamaterial, where $\omega$ is the angular frequency and $\mathbf{k}=(k_x,k_y,k_z)$ is the wave vector. The possible dependence of the dielectric function $\overline{\overline{\varepsilon}}_{eff}(\omega,\mathbf{k})$ on the wave vector results from spatial dispersion effects [26], which are characteristic of several metamaterials, even for very low frequencies [5, 12, 22, 27].

In order to compute the unknown dielectric function for a given $(\omega,\mathbf{k})$, the composite material is excited with a Floquet-periodic external distribution of electric current $\mathbf{J}_e$ of the form $\mathbf{J}_e = \mathbf{J}_{e,\mathbf{av}} e^{-j\mathbf{k}\cdot\mathbf{r}}$ where $\mathbf{J}_{e,\mathbf{av}}$ is a constant vector. Hence, the induced "microscopic" electric and induction fields ($\mathbf{E}$, $\mathbf{B}$) have also the Floquet property, and verify the "microscopic" Maxwell equations:

$$\nabla \times \mathbf{E} = -j\omega \mathbf{B} \tag{1a}$$

$$\nabla \times \frac{\mathbf{B}}{\mu_0} = \mathbf{J}_e + \mathbf{J}_d + j\varepsilon_0 \omega \mathbf{E}, \tag{1b}$$

where $\mathbf{J}_d = \varepsilon_0(\varepsilon_r - 1)j\omega\mathbf{E}$ is the induced current relative to the host medium and $\varepsilon_r$ is the periodic permittivity of the material.

For a two-dimensional material, the macroscopic average macroscopic fields $\mathbf{E}_{av}$ and $\mathbf{B}_{av}$ are defined as follows:

$$\mathbf{E}_{av} = \frac{1}{A_{cell}} \int_\Omega \mathbf{E}(\mathbf{r}) e^{+j\mathbf{k}\cdot\mathbf{r}} d^2\mathbf{r}, \qquad \mathbf{B}_{av} = \frac{1}{A_{cell}} \int_\Omega \mathbf{B}(\mathbf{r}) e^{+j\mathbf{k}\cdot\mathbf{r}} d^2\mathbf{r}. \qquad (2)$$

In the above, $\Omega$ is the unit cell of the material and $A_{cell} = |\mathbf{a}_1 \times \mathbf{a}_2|$ is the area of $\Omega$. It can be easily verified that the macroscopic fields verify the following equations:

$$-\mathbf{k} \times \mathbf{E}_{av} + \omega \mathbf{B}_{av} = 0 \qquad (3a)$$

$$\omega \mathbf{D}_{g,av} + \mathbf{k} \times \frac{\mathbf{B}_{av}}{\mu_0} = -\omega \mathbf{P}_e, \qquad (3b)$$

where $\mathbf{P}_e = \frac{1}{j\omega A_{cell}} \int_\Omega \mathbf{J}_e(\mathbf{r}) e^{+j\mathbf{k}\cdot\mathbf{r}} d^2\mathbf{r}$ is by definition the applied polarization vector and the generalized electric displacement $\mathbf{D}_{g,av}$ verifies the constitutive relation,

$$\mathbf{D}_{g,av} \equiv \varepsilon_0 \mathbf{E}_{av} + \mathbf{P}_g = \overline{\overline{\varepsilon}}_{eff}(\omega, \mathbf{k}) \cdot \mathbf{E}_{av}. \qquad (4)$$

In Eq. (4), $\mathbf{P}_g = \frac{1}{j\omega A_{cell}} \int_\Omega \mathbf{J}_d(\mathbf{r}) e^{+j\mathbf{k}\cdot\mathbf{r}} d^2\mathbf{r}$ is the so-called generalized polarization vector, which can be related to the classical polarization and magnetization vectors and with higher-order multipoles [20]. It should be clear from Eq. (4) that the dielectric function can be determined for a given $(\omega, \mathbf{k})$ provided $\mathbf{P}_g$ is known for three independent vectors $\mathbf{E}_{av}$ (e.g., for $\mathbf{E}_{av} \parallel \hat{\mathbf{u}}_i$, where $\hat{\mathbf{u}}_i$ is directed along the coordinate axes). Hence, the unknown dielectric function can be computed using the procedure outlined next.

The first step is to select three independent vector amplitudes $\mathbf{J}_{e,av}$ for the applied current $\mathbf{J}_e$ ($\omega$ and $\mathbf{k}$ are fixed), and, for each distribution of current, solve the Maxwell's Equations [Eq. (1)] to obtain the microscopic fields. It is assumed that the obtained average fields $\mathbf{E}_{av}$ form an independent set of vectors. In the second step, the generalized polarization vector $\mathbf{P}_g$ associated with each distribution of microscopic fields is determined using the formula $\mathbf{P}_g = \frac{1}{j\omega A_{cell}} \int_\Omega \mathbf{J}_d(\mathbf{r}) e^{+j\mathbf{k}\cdot\mathbf{r}} d^2\mathbf{r}$, and then finally the dielectric function of the material $\bar{\bar{\varepsilon}}_{eff} = \bar{\bar{\varepsilon}}_{eff}(\omega, \mathbf{k})$ is calculated so that it is consistent with Eq. (4). The calculated dielectric function is independent of the excitation, i.e. it is independent of the specific set of constant vectors $\mathbf{J}_{e,av}$ that is considered in the calculation.

An important property discussed in Ref. [20] (see also Refs. [28, 29]) is that in presence of weak spatial dispersion, so that the material can be described using conventional constitutive relations and its response is to some degree of approximation local, the effective parameters of the artificial medium (local permittivity, local permeability, and magnetoelectric parameters) can be readily extracted from the nonlocal dielectric function. Specifically, the relation between the nonlocal dielectric function and the local parameters is as follows:

$$\frac{\bar{\bar{\varepsilon}}_{eff}}{\varepsilon_0}(\omega, \mathbf{k}) = \bar{\bar{\varepsilon}}_r - \bar{\bar{\xi}} \cdot \bar{\bar{\mu}}_r^{-1} \cdot \bar{\bar{\zeta}} + \frac{c}{\omega}\left(\bar{\bar{\xi}} \cdot \bar{\bar{\mu}}_r^{-1} \times \mathbf{k} - \mathbf{k} \times \bar{\bar{\mu}}_r^{-1} \cdot \bar{\bar{\zeta}}\right) + \frac{c^2}{\omega^2} \mathbf{k} \times \left(\bar{\bar{\mu}}_r^{-1} - \bar{\bar{\mathbf{I}}}\right) \times \mathbf{k} \quad (5)$$

where $\bar{\bar{\mathbf{I}}}$ is the identity dyadic, $\bar{\bar{\varepsilon}}_r(\omega)$ is the *local* permittivity dyadic (relative to the free-space permittivity), $\bar{\bar{\mu}}_r(\omega)$ is the *local* permeability (relative to the free-space permeability), and $\bar{\bar{\xi}}(\omega)$ and $\bar{\bar{\zeta}}(\omega)$ are dimensionless tensors that characterize the magneto-electric coupling [20]. The constitutive relations implicit in the definition of the local parameters are [20]:

$$\mathbf{D} = \varepsilon_0 \overline{\overline{\varepsilon_r}} . \mathbf{E}_{av} + \sqrt{\varepsilon_0 \mu_0} \, \overline{\overline{\xi}} . \mathbf{H} \tag{6a}$$

$$\mathbf{B}_{av} = \sqrt{\varepsilon_0 \mu_0} \, \overline{\overline{\zeta}} . \mathbf{E}_{av} + \mu_0 \overline{\overline{\mu_r}} . \mathbf{H} \tag{6b}$$

It should be clear that the above constitutive relations are different from the constitutive relation implicit in Eq. (4). The main advantage of the constitutive relations (6) is that they are local, and in particular the number of parameters that characterize the material is drastically reduced (because the local parameters are independent of the wave vector). Moreover, the constitutive relations (6) are valid both in the spectral domain as well as in the spatial domain, and this simplifies considerably the analysis of problems involving interfaces between different media. The constitutive relations (6) have been used for a long time in the characterization of media with optical activity [29, 30, 31]. As mentioned before, in this work it is assumed that the geometry of the metamaterial is intrinsically two-dimensional (axis of the structure is along $z$) and that $k_z = 0$. Moreover, it is also supposed that the electromagnetic wave is transverse electric to $z$ (TE$^z$). In such conditions it may be assumed that $\overline{\overline{\mu_r}}(\omega) = \mu_{r,zz} \hat{\mathbf{u}}_z \hat{\mathbf{u}}_z$, and the relative magnetic permeability of the medium $\mu_{r,zz}$ can be numerically calculated as follows [20]:

$$\frac{\mu_{eff}}{\mu_0} \equiv \mu_{r,zz}(\omega) = \frac{1}{1 - \left(\dfrac{\omega}{c}\right)^2 \dfrac{1}{2\varepsilon_0} \dfrac{\partial^2 \varepsilon_{eff,yy}}{\partial k_x^2}\bigg|_{\mathbf{k}=0}}, \tag{7}$$

where $\varepsilon_{eff,yy} = \hat{\mathbf{u}}_y \cdot \overline{\overline{\varepsilon}}_{eff} \cdot \hat{\mathbf{u}}_y$. Formula (7) shows that the emergence of artificial magnetism is intrinsically related to spatial dispersion effects of second order [20, 28]. On the other end, putting $\mathbf{k} = 0$ in Eq. (5) it is readily found that:

$$\frac{\overline{\overline{\varepsilon_{eff}}}}{\varepsilon_0}(\omega, \mathbf{k} = 0) = \overline{\overline{\varepsilon_r}} - \overline{\overline{\xi}} . \overline{\overline{\mu_r}}^{-1} . \overline{\overline{\zeta}}. \tag{8}$$

In the absence of magneto-electric coupling $\bar{\bar{\xi}} = \bar{\bar{\zeta}} = 0$ (e.g. if the material has inversion symmetry, i.e. it is invariant under the transformation $\mathbf{r} \to -\mathbf{r}$ with respect to some suitable origin of the coordinates), the local electric permittivity is given by $\bar{\bar{\varepsilon}}(\omega) = \bar{\bar{\varepsilon}}_{eff}(\omega, \mathbf{k} = 0)$. However, in general the tensors $\bar{\bar{\xi}}$ and $\bar{\bar{\zeta}}$ do not vanish. The relevant components of the tensor $\bar{\bar{\zeta}}$ for the two-dimensional geometry under study (TE$^z$–polarized waves with $k_z = 0$) are clearly $\zeta_{zx}$ and $\zeta_{zy}$. Thus, we may assume that

$$\bar{\bar{\zeta}} = \zeta_{zx}\hat{\mathbf{u}}_z\hat{\mathbf{u}}_x + \zeta_{zy}\hat{\mathbf{u}}_z\hat{\mathbf{u}}_y \tag{9}$$

where $\hat{\mathbf{u}}_z\hat{\mathbf{u}}_x \equiv \hat{\mathbf{u}}_z \otimes \hat{\mathbf{u}}_x$ represents the tensor product of two vectors. Taking into account that in reciprocal media the tensor $\bar{\bar{\xi}}$ is linked to $\bar{\bar{\zeta}}$ by the relation $\bar{\bar{\xi}} = -\bar{\bar{\zeta}}^T$ [31] (the superscript "$T$" represents the transpose tensor), and substituting Eq. (9) into Eq. (5), it is found after straightforward calculations that

$$\zeta_{zx} = -\frac{\omega}{c}\mu_{r,zz}\frac{1}{\varepsilon_0}\frac{\partial \varepsilon_{eff,xy}}{\partial k_x}\bigg|_{\mathbf{k}=0}, \qquad \zeta_{zy} = -\frac{\omega}{c}\mu_{r,zz}\frac{1}{\varepsilon_0}\frac{\partial \varepsilon_{eff,xy}}{\partial k_y}\bigg|_{\mathbf{k}=0} \tag{10}$$

Hence, the magneto-electric parameters are obtained from the first order derivatives of the nonlocal dielectric function with respect to the wave vector.

### B. 2-D FDFD method

In order to determine the "microscopic" electric fields [solution of Eq. (1)], we employ the well-known FDFD method based on the Yee's mesh [32]. The finite differences (FD) method is excellent to model devices with a complex geometry or structures of finite size. It is an accurate and stable method where the sources of error such as the grid resolution, nonphysical reflections from the grid boundaries, and the effect of representing curved surfaces on a Cartesian grid, are well understood. Being a frequency-domain method, it is able to resolve sharp resonances and obtain solutions at a single frequency more efficiently than time-domain methods. In the FD method the unit cell Ω is divided into many rectangular grids. A portion of the grid with a dielectric inclusion is illustrated in Fig. 1.

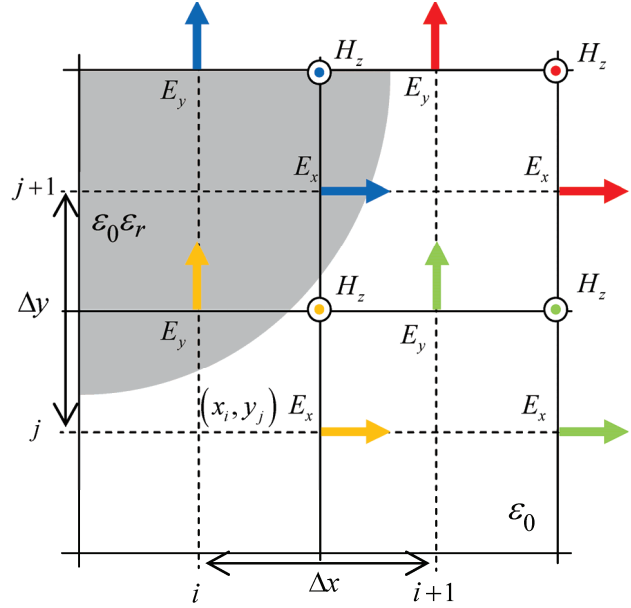

Fig. 1. Geometry of the grid mesh for the FDFD method. The nodes are spaced by $\Delta x$ and $\Delta y$ along the $x$- and $y$-directions, respectively. The shaded region represents a portion of the dielectric inclusion with dielectric permittivity $\varepsilon_r$.

For TE$^{-z}$ polarized waves the electromagnetic field only has the following Cartesian components $E_x, E_y, H_z$, and the wave vector is restricted to the $xoy$ plane ($k_z = 0$). Hence, in our case, the problem to be solved [Eq. (1)] is:

$$\begin{cases} \dfrac{\partial^2 E_y}{\partial x \partial y} - \dfrac{\partial^2 E_x}{\partial_y^2} - \left(\dfrac{\omega}{c}\right)^2 \varepsilon_r E_x = -j\omega\mu_0 J_{e,x} \\ \dfrac{\partial^2 E_x}{\partial x \partial y} - \dfrac{\partial^2 E_y}{\partial_x^2} - \left(\dfrac{\omega}{c}\right)^2 \varepsilon_r E_y = -j\omega\mu_0 J_{e,y} \end{cases}. \quad (11)$$

In the above, $J_{e,x}$ and $J_{e,y}$ are the components of the applied current density ($\mathbf{J}_e = \mathbf{J}_{e,\mathrm{av}} e^{-j\mathbf{k}\cdot\mathbf{r}}$) along the $x$ and $y$ coordinates. In order to discretize the derivatives of the electric fields in Eq. (11) we employ the formulas proposed in [33]:

$$\frac{\partial^2 E_y}{\partial x^2}(i,j) = \frac{E_y(i+1,j) - 2E_y(i,j) + E_y(i-1,j)}{(\Delta x)^2} \tag{12a}$$

$$\frac{\partial^2 E_y}{\partial x \partial y}(i,j) = \frac{E_y(i+1,j) - E_y(i,j) - E_y(i+1,j-1) + E_y(i,j-1)}{\Delta x \Delta y} \tag{12b}$$

$$\frac{\partial^2 E_x}{\partial y^2}(i,j) = \frac{E_x(i,j+1) - 2E_x(i,j) + E_x(i,j-1)}{(\Delta y)^2} \tag{12c}$$

$$\frac{\partial^2 E_x}{\partial x \partial y}(i,j) = \frac{E_x(i,j+1) - E_x(i,j) - E_x(i-1,j+1) + E_x(i-1,j)}{\Delta x \Delta y}, \tag{12d}$$

where $\Delta x$ and $\Delta y$ is the grid spacing along the $x$- and $y$-directions, respectively. The discrete indices $(i,j)$ attached to the field components are such that $E_x(i,j) \equiv E_x|_{(x_i + \Delta x/2, y_j)}$, $E_y(i,j) \equiv E_y|_{(x_i, y_j + \Delta y/2)}$, where $(x_i, y_j)$ are the Cartesian coordinates of the considered node (see Fig. 1). If the mesh of Fig. 1 has $N$ nodes in the unit cell, then there are $2N$ unknowns, as each node corresponds to two components of the electric field, $E_x$ and $E_y$ respectively. The magnetic field $H_z = -\frac{1}{j\omega\mu_0}\left(\frac{\partial E_y}{\partial x} - \frac{\partial E_x}{\partial y}\right)$ can be calculated at each node using the formula [24],

$$H_z(i,j) = -\frac{1}{j\omega\mu_0}\left(\frac{E_y(i+1,j) - E_y(i,j)}{\Delta x} - \frac{E_x(i,j+1) - E_x(i,j)}{\Delta y}\right), \tag{13}$$

being $H_z(i,j) \equiv H_z|_{(x_i + \Delta x/2, y_j + \Delta y/2)}$. All the nodes situated at the boundary of the unit cell have some adjacent nodes lying outside of the unit cell, but they can be "brought back" using the Bloch-Floquet periodic boundary conditions,

$$\Phi(x+a, y+b) = e^{-jk_x a - jk_y b}\Phi(x,y) \tag{14}$$

Here, $\Phi(x,y)$ is any field component, $k_x$ and $k_y$ are the wave vector components along the $x$- and $y$- directions, and $a$ and $b$ are the lattice constants along the $x$- and $y$- directions. Substituting Eqs. (12) into the system (11) and taking into account the Bloch-Floquet boundary conditions (14), it is possible to reduce the homogenization problem to a standard linear system that can be numerically solved with respect to the unknowns (microscopic electric field components at the grid nodes). In this manner, we obtain the microscopic fields, which are then used to compute the nonlocal dielectric function as detailed in Sec II A. Since the problem is effectively two-dimensional we only need to solve two excitation problems to retrieve the nonlocal dielectric function.

### III. APPLICATION OF THE HOMOGENIZATION METHOD

#### A. Dielectric cylinders

To validate the developed numerical code, first we computed the effective dielectric function of a metamaterial formed by cylindrical dielectric inclusions with circular cross-section and dielectric constant $\varepsilon_r = \varepsilon_r' - j\varepsilon_r''$ with $\varepsilon_r' \geq 1$. The inclusions are arranged in a square lattice with period $a$ and have radius $R$ (Fig 2). The real part of the effective dielectric function (Fig. 2) was computed for $\mathbf{k} = 0$ and for the normalized frequency $\omega a / c = 0.001$ (quasi-static regime). A uniform mesh was used with $\Delta x = \Delta y = a/34$ and the computation time for each frequency sample is approximately 15s in a standard personal computer [34]. The calculated effective permittivity was compared with the effective permittivity extracted from the band structure of the periodic material (slope of the fundamental mode near the origin of the Brillouin zone) using the hybrid plane-wave integral-equation-based method described in Refs. [35-36].

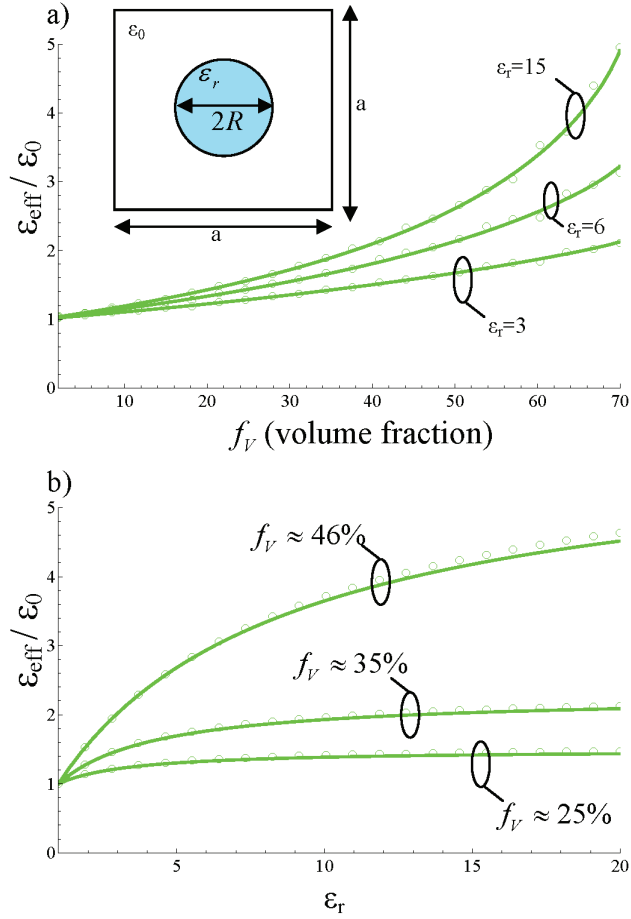

Fig. 2. (Color online) (a): Quasi-static effective permittivity as a function of the volume fraction $f_V$ (in percentage) of the inclusions for different values of the permittivity $\varepsilon_r$. (b): effective permittivity as a function of $\varepsilon_r$ for different values of $f_V$. The discrete symbols in (a) and (b) correspond to the values computed with the homogenization method and the solid lines were obtained from the slope of the band structure of the material at the origin of the Brillouin zone. The geometry of the unit cell of the two dimensional metamaterial is shown in the inset: the unit cell consists of a cylindrical inclusion with circular cross-section, normalized radius $R/a$, and permittivity $\varepsilon_r$.

In Fig. 2a the computed results are shown as a function of the volume fraction of the cylinders (in percentage) for different values of the permittivity: $\varepsilon_r = 3$, $\varepsilon_r = 6$ and $\varepsilon_r = 15$ (the metamaterial is assumed lossless). It can be seen that the results obtained using the homogenization method concur very well with the results obtained from the slope of the band structure at the origin of the Brillouin. As expected, the effective permittivity grows with the volume fraction of the cylinders. In Fig. 2b the effective permittivity is plotted as a function of the relative permittivity of the cylinders, now for a fixed volume fraction of the inclusions. It is seen that the effective permittivity is more sensitive to the variation of $\varepsilon_r$ for cylinders with large radii.

In a second example we have extracted both $\varepsilon_{eff}$ and $\mu_{eff}$ for a mixture of high-index cylinders with permittivity $\varepsilon_r = 56$ and normalized radius $R = 0.4a$ (Fig 3). Metamaterials formed by a mixture of cylinders with large dielectric constant may exhibit a strong magnetic response due to the excitation of the Mie resonance in the particles [37]. Inclusions with a large dielectric constant are used so that free-space wavelength is much larger than the spacing between inclusions at the resonance frequency, and consequently the system may be considered as an effective medium. Related configurations (with spherical inclusions) have been studied in Refs. [38,39] to mimic the response of an isotropic double-negative medium [38,39].

Using the FDFD implementation of the nonlocal homogenization method, we computed the effective parameters of the considered system as a function of the normalized frequency $\omega a / c$ (Fig 3). Consistent with Ref. [37], it can be seen that the composite material may have an effective response very different from its non-magnetic dielectric constituents, and in particular, that the magnetic response may be greatly enhanced around $\omega a / c \approx 0.8$, while the electric response has a resonance near $\omega a / c \approx 1.2$.

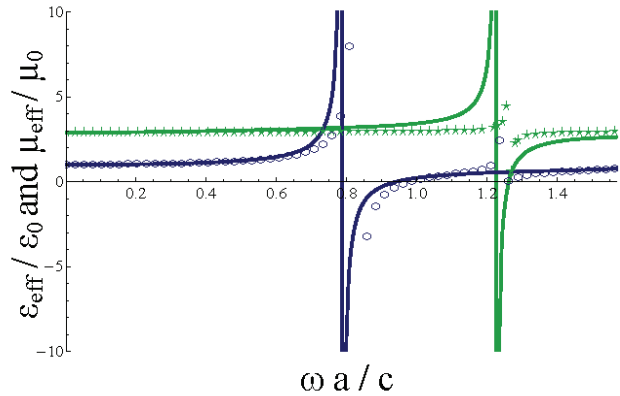

Fig. 3.(Color online) Real parts of the effective permittivity $\varepsilon_{eff}$ (green curves) and permeability $\mu_{eff}$ (blue curves) as a function of the normalized frequency $\omega a / c$. The discrete symbols correspond to the values extracted with the FDFD method and the solid lines are the values obtained using the Clausius-

Mossotti formula. The high-index cylindrical-shaped inclusion has a normalized radius $R = 0.4a$ and permittivity $\varepsilon_r = 56$. The host material is air.

We compared the results obtained using the homogenization method with those predicted by the Clausius-Mossotti mixing formula (the expressions of the dynamic electric and magnetic polarizabilities of the dielectric cylinder can be found in Ref. [40]). As can be seen in Fig. 3, despite the relatively large diameter of the cylinders, the results extracted with the FDFD method agree surprisingly well with those obtained using the Clausius-Mossotti formula. This indicates that the interaction between the inclusions is predominantly of the dipole-type.

### B. Plasmonic inclusions

Particles with a plasmonic response are important building blocks of optical metamaterials. Since the early pioneering works of Bergman and Milton it is well-known that nanoparticles with negative permittivity can support multiple electric resonances [41, 42]. Several configurations that exploit the existence of such plasmonic resonances in order to tailor the effective properties of a composite material (notably the magnetic response in the optical regime) have been put forward in the recent literature [43-46]. It is thus relevant to characterize the optical response of a metamaterial formed by an array of such particles. For simplicity, it is supposed here that the permittivity of the inclusions is described by the Drude model $\varepsilon_r = 1 - \dfrac{\omega_p^2}{\omega(\omega - j\Gamma)}$, where $\omega_p$ is the plasma frequency and $\Gamma$ is the collision frequency. The Drude dispersion model may describe accurately the response of noble metals through the infrared and optical domains. It is assumed that the normalized plasma frequency verifies $\omega_p a / c = 1$, where $a$ is the lattice period. The inclusions are arranged in a square lattice and have a normalized radius $R = 0.45a$. These parameters are the same as in Ref. [43].

The extracted permittivity ($\varepsilon_{eff} = \varepsilon_{eff}(\omega, \mathbf{k} = 0)$) is depicted in Fig. 4 as a function of the normalized frequency $\omega/\omega_p$. Clearly, the effective permittivity of the metamaterial has several sharp singularities, especially close to the frequency $\omega/\omega_p = 0.7$, which corresponds to the plasmon resonance for a single cylindrical particle ($\varepsilon = -1$). This irregular behavior of the electric response is a consequence of the excitation of multiple quasi-static resonances that are characteristic of closely coupled plasmonic particles, consistent also with the results of Ref. [43], which showed that in general the material may support almost dispersionless bulk plasmons, propagating plasmon polaritons, and modes associated with high-multipole resonances. It Ref. [43] it was shown that these "high-multipole resonances" may be associated with a regime where the structure behaves as a double negative material. This is indeed supported by our numerical analysis, which shows that in the vicinity of $\omega/\omega_p = 0.63$ both the effective permittivity and the effective permeability are simultaneously negative [the effective permeability is shown in the inset of Fig. 4, and was evaluated using Eq. (7)]. In particular, we have obtained that at $\omega/\omega_p \approx 0.637$ the effective permittivity is $\varepsilon_{eff} = -0.56$, and the effective permeability is $\mu_{eff} = -2.35$, consistent with the values reported in Ref. [43] for a nearby frequency ($\mu_{eff} = -2.35$ and $\varepsilon_{eff} = -0.427$ at $\omega/\omega_p = 0.6$).

When the absorption is increased (i.e. $\Gamma$ is increased), so that the surface plasmon polaritons are more damped, the somehow irregular behavior of the electric response tends to disappear. The effect of loss is particularly important at the frequencies associated with the plasmonic resonances.

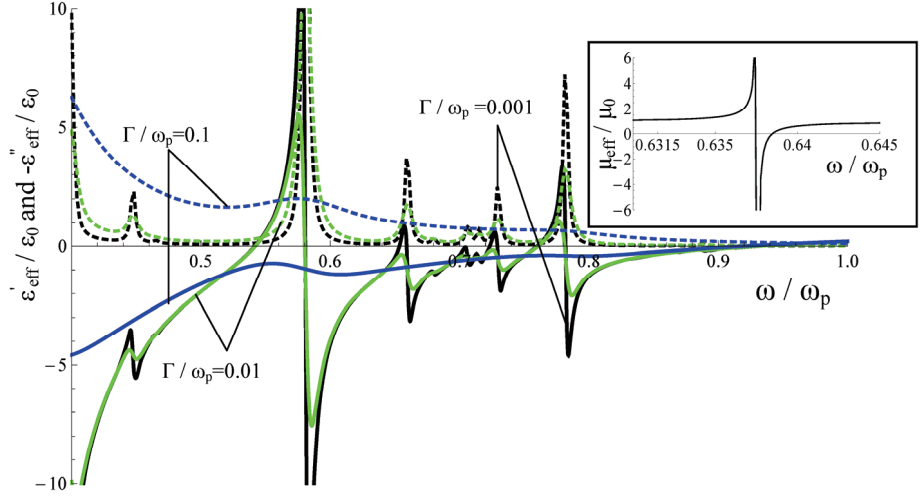

Fig. 4. (Color online) Real and imaginary parts of the effective permittivity $\varepsilon_{eff}/\varepsilon_0 = \varepsilon' - j\varepsilon''$ as a function of the normalized frequency $\omega/\omega_p$, for a mixture with plasmonic-cylinders arranged in a regular lattice, for different values of the damping frequency: $\Gamma/\omega_p = 0.001$ (black curves), $\Gamma/\omega_p = 0.01$ (green curves) and $\Gamma/\omega_p = 0.1$ (blue curves). The solid lines correspond to the real part of $\varepsilon_{eff}$ while the dashed lines represent the imaginary part. The host material is air. The inset shows the real part of the effective magnetic permeability ($\Gamma/\omega_p = 0.001$) of the system in the vicinity of $\omega/\omega_p = 0.64$.

To further investigate the properties of the plasmonic resonances and emphasize their quasi-static nature [47], we have studied a second metamaterial configuration where the inclusions are square-shaped tilted plasmonic inclusions (with negative permittivity), arranged in a square lattice (inset of Fig. 5) [48]. The computed permittivity is shown in Fig. 5 as a function of the (negative) permittivity of the square-shaped inclusion. The quasi-static regime is assumed ($\omega \approx 0$) and the square cylinders have sharp corners. As shown in Fig. 5, somehow similar to the previous example, the effective permittivity of the metamaterial has several singularities, which are a consequence of the excitation of multiple quasi-static plasmonic resonances. However, in the present case the resonant response is enhanced by the fact that the boundary of the inclusions is not a smooth curve. More specifically, as also discussed in Ref. [48], the sharp corners of the plasmonic inclusion may cause a great enhancement of the electromagnetic field in their vicinity. Thus, unlike the array of circular cylinders considered in the previous example (for which the inclusions have a smooth boundary), the array of square cylinders supports strongly localized resonances which stem specifically from the irregular boundary of the inclusions. We compared the results obtained with our method with the effective permittivity calculated in Ref. [48] using a different method (discrete symbols in Fig. 5). The general agreement between both sets of results is quite satisfactory, but it may be noticed that our extraction method does not predict as many singularities as the method used in Ref. [48]. The reason may be that in Ref. [48] a refined mesh was used at the sharp corners while our numerical code uses a uniform mesh, hence predicting less accurately the singularities resultant from the sharp corners (whose number is infinite in the range $-3 < \varepsilon_r < -1/3$).

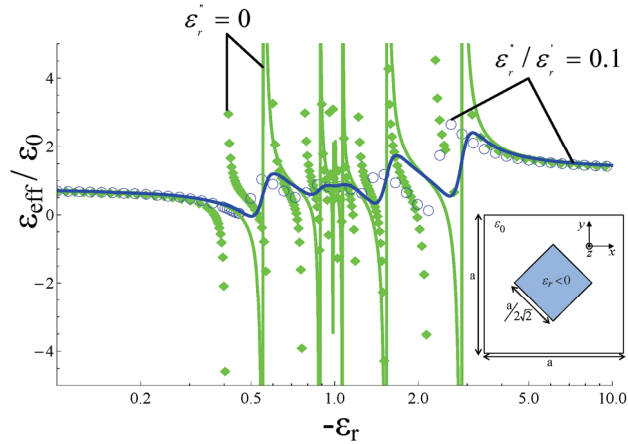

Fig. 5. (Color online) Real part of the effective quasi-static permittivity as a function of the real part of the permittivity of the inclusion $\varepsilon'_r$, with: $\varepsilon''_r = 0$ (solid green curve and discrete diamond-shaped green symbols), $-\varepsilon''_r/\varepsilon'_r = 0.1$ (solid blue curve and discrete circle-shaped blue symbols). The geometry of the unit cell of the two dimensional metamaterial is shown in the inset: the unit cell consists of a square-shaped inclusion with negative permittivity $\varepsilon_r = \varepsilon'_r - j\varepsilon''_r$ and sharp corners. The solid curves were extracted with the FDFD method while the discrete symbols are from Ref. [48]. The host material is air.

It was shown in Ref. [48] that when the inclusion has negative permittivity and an infinitely sharp wedge, the electromagnetic fields may have a very irregular behavior at the wedge and the stored energy may not be finite. These problems can be minimized by adding loss to the material. Indeed, it may be seen in Fig. 5 that when losses are considered, i.e., the permittivity of the inclusions is of the form $\varepsilon_r = \varepsilon'_r - j\varepsilon''_r$ with $\varepsilon''_r > 0$, the amplitude of the resonances is damped. However, it should be mentioned that even the presence of absorption may not be sufficient to avoid a resonant response. For more details about this counterintuitive result the reader is referred to Ref. [48], where it is shown that to obtain a stable response it is necessary to add both loss and round the corners of the inclusions.

### C. Zero-index media

Zero-index media is a class of metamaterials with index of refraction equal to zero (or near zero) at the frequency of operation. Due to the relatively long wavelengths intrinsic to these materials, they may have interesting potentials in tunneling electromagnetic energy through narrow channels and bends [3, 4, 40], to increase the directivity of an antenna [49], to manipulate the shape of wave fronts, and to design delay lines [50]. In general, zero-index media are strongly mismatched with free-space due to the huge difference between the wave impedance $\eta = \sqrt{\mu/\varepsilon}$ in such materials and free-space. Nevertheless, when both the permeability and the permittivity of the material are near zero, the metamaterial may have near zero index and may be matched to free-space.

Clearly, the realization of a material with such properties is not a trivial matter. One of the possible realizations is a racemic mixture of left-handed and right-handed helices of certain pitch angle, as considered in [51]. In a recent work [40], a different strategy to obtain a material with simultaneously near zero permittivity and permeability was explored. It was shown that such material may be easily realized provided a material with near zero permittivity (and no magnetic response) is somehow available. This may be the case of some metals at optical and UV frequencies, and some semiconductors and polar dielectrics at infrared frequencies. Specifically, it was shown in Ref. [40] that by embedding dielectric particles with suitable size and permittivity in a host background with near zero permittivity it may be possible to realize a composite material with simultaneously near zero permittivity and permeability. Our objective here is to characterize the effective parameters of such zero-index composite materials.

For simplicity, we suppose that the geometry of the metamaterial is similar to that of Sec. III.A. Specifically, the metamaterial is formed by cylindrical-shaped inclusions with normalized radius $R/a = 0.4$, and arranged in a square lattice. However, instead of being embedded in air, we will consider that the cylindrical inclusions are embedded in an epsilon-near-zero (ENZ) host medium ($\varepsilon_h \approx 0$). It was demonstrated in Ref. [40] that at the frequency where $\varepsilon_h = 0$, the effective permittivity of the composite material vanishes, whereas the effective permeability is given by the following (exact) formula:

$$\mu_{eff} = \mu_0 \left( \frac{A_{h,cell}}{A_{cell}} + \frac{2\pi R^2}{A_{cell}} \frac{1}{k_r R} \frac{J_1(k_r R)}{J_0(k_r R)} \right), \qquad \text{at } \omega = \omega_p \qquad (15)$$

where $A_{h,cell} = A_{cell} - \pi R^2$, $k_r = \omega\sqrt{\varepsilon_r \varepsilon_0 \mu_0}$ and $J_l$ is the Bessel function of 1$^{st}$ kind and order $l$. In order to further validate our numerical code, we have computed the effective permeability of the composite material using the FDFD homogenization method. The effective permeability is written in terms of the derivatives of the nonlocal dielectric function with respect to the wave vector [Eq. (7)]. The calculated result is depicted in Fig. 6 (discrete symbols) as a function of the relative dielectric permittivity of the cylinders $\varepsilon_r$, and compares very well with exact result given by Eq. (15) (solid lines). It may be seen that the magnetic response has a resonance when $\varepsilon_r \approx 36$, and that the permeability is near zero, $\mu_{eff} \approx 0$, when $\varepsilon_r \approx 56$.

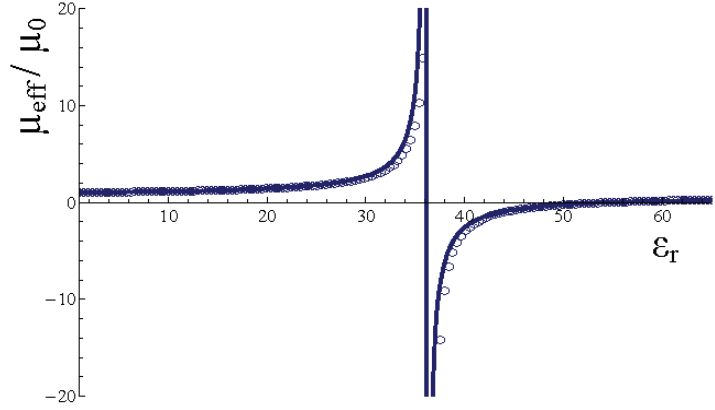

Fig. 6. (Color online) Real part of the effective permeability $\mu_{eff}$ as a function of the permittivity $\varepsilon_r$ of a square lattice of dielectric cylinders embedded in a host medium with permittivity near to zero ($\varepsilon_h \approx 0$). The cylindrical-shaped inclusions have normalized radius $R = 0.4a$. The solid curve was obtained using Eq. (15) and the discrete symbols where obtained with the FDFD method.

In order to study the frequency response of the metamaterial we have calculated the local permittivity and the local permeability as functions of frequency. It is assumed that the host permittivity is described by the Drude dispersion model, and that the permittivity of the cylinder is $\varepsilon_r = 56$. The normalized plasma frequency verifies $\omega_p a/c = 1.0$ and the collision frequency is such that $\Gamma/\omega_p = 0.001$.

In Fig. 7a the extracted effective parameters (discrete symbols) are depicted as function of the frequency, showing that, consistent with the theory of [40], both the permittivity and permeability are near zero at the plasma frequency: $\varepsilon_{eff}(\omega_p) = \mu_{eff}(\omega_p) \approx 0$. The solid lines in Fig. 7a represent the effective parameters predicted by the Clausius-Mossotti mixing formula (see Ref. [40]), being the general agreement with the data extracted with the FDFD homogenization quite good. In Fig. 7b we show the amplitude of the electric field component $E_y$ at different frequencies of operation marked in Fig. 7a, and supposing that the excitation $\mathbf{J}_e = \mathbf{J}_{e,av} e^{-j\mathbf{k}\cdot\mathbf{r}}$ is directed along $y$ and $\mathbf{k} = 0$. It may be seen in Fig. 7b (i) that when $\omega/\omega_p \approx 1$ ($\varepsilon_h \approx 0$) the electric field does not penetrate into the cylinder and is strongly concentrated in the ENZ region, so that the spatially average electric displacement vector $\langle D_y \rangle = \langle \varepsilon E_y \rangle$ vanishes ($\varepsilon_{eff} \approx 0$). The situation is quite different at the resonance of the electric response, $\omega/\omega_p \approx 1.282$, as shown in Fig. 7b (ii).

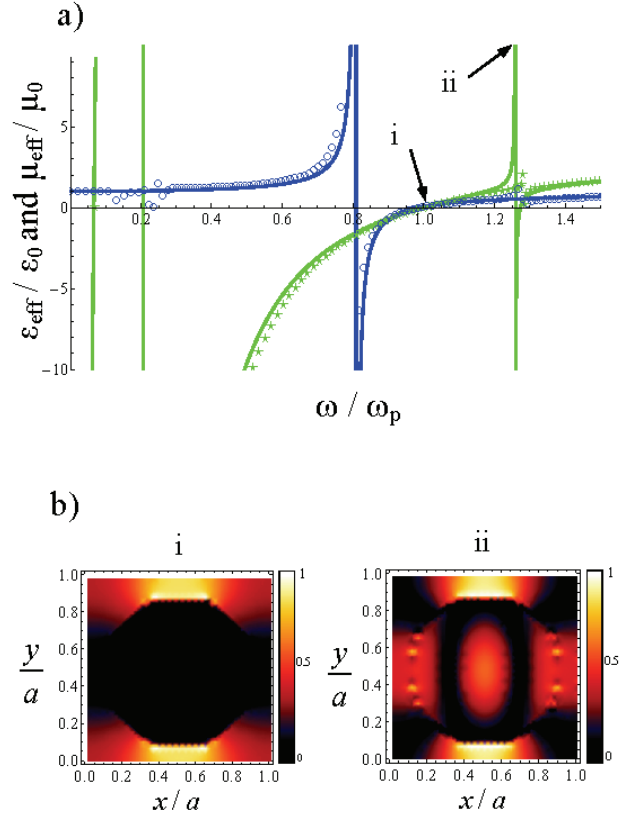

Fig. 7.(Color online) (a): Real parts of the effective permittivity $\varepsilon_{eff}$ (green curves) and permeability $\mu_{eff}$ (blue curves) as a function of the normalized frequency $\omega/\omega_p$. The discrete symbols correspond to the values extracted with the FDFD method and the solid lines were obtained using the Clausius-Mossotti formula. The inclusion has a normalized radius $R = 0.4a$ and permittivity $\varepsilon_r = 56$ and the host medium is characterized by a Drude type dispersion model. (b) normalized amplitude of the electric field component $E_y$ in the unit cell (when the current source is directed along $y$) at the frequency where the effective permittivity hits a resonance (ii) and when $\varepsilon_{eff} \approx 0$ (i).

### D. Horseshoe inclusion

An interesting proposal to obtain a strong magnetic response using plasmonic nano-particles is based on nanostructures shaped as a horseshoe. It was shown in Ref. [52] that such "nanoantennas" with dimensions much smaller than the light wavelength can have a magnetic plasmon resonance with resonant frequency depending on the shape and material properties rather than on the wavelength. In this section, we will study the effective magnetic response of arrays of such metallic horseshoe-shaped nanostructures. The two-dimensional metamaterial consists of metallic nanoantennas arranged in a square lattice (Fig. 8a). The permittivity $\varepsilon_r$ of the inclusions follows the Drude dispersion model, being the normalized plasma frequency $\omega_p a/c = 30.0$ and the collision frequency $\Gamma/\omega_p = 0.001$. The normalized thickness of the "arms" of the nanoantennas is $b/a = 0.18$ and the distance between the opposite arms is $d/a = 0.26$. Using the homogenization FDFD method we calculated the effective permeability $\mu_{eff}$ of the composite material as a function of the normalized frequency $\omega a/c$ (solid line Fig. 8a) (as before the effective permeability is calculated from the second order derivatives of the nonlocal dielectric function with respect to the wave vector). The homogenization method predicts that there is a magnetic resonance around $\omega a/c = 1.47$.

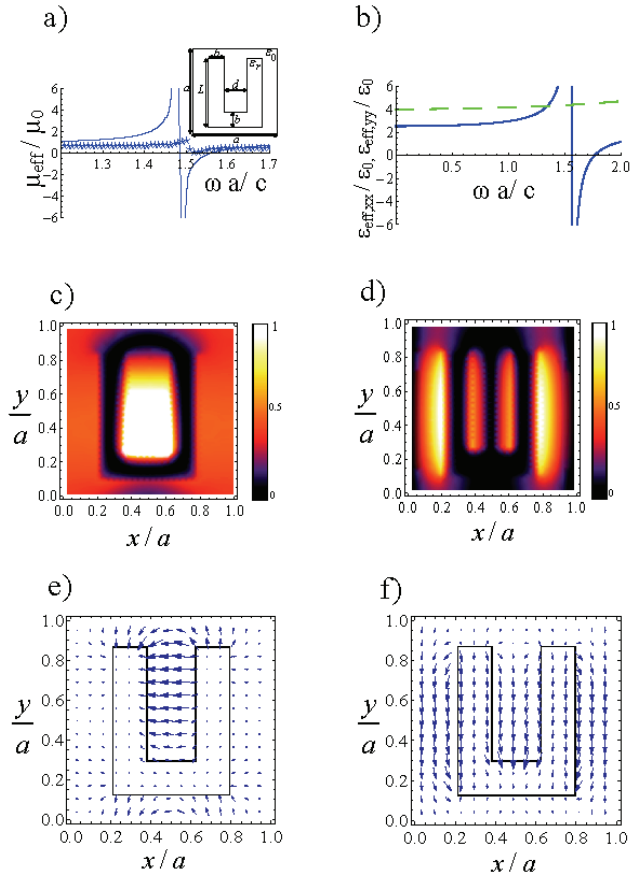

Fig. 8. (Color online) (a): Real part of the effective permittivity $\mu_{eff}$ as a function of the normalized frequency $\omega a/c$. Solid line: homogenization method used in this work; Discrete symbols: data extracted using the inversion of the reflection and transmission coefficients [16-17]. The geometry of the unit cell is shown in the inset and consists of a U-shaped plasmonic inclusion with complex permittivity $\varepsilon_r$. The arms of the horseshoe have a normalized thickness $b = 0.18a$ and normalized length $L = 0.79a$. The distance between the two arms is $d = 0.26a$. (b) Real parts of $\varepsilon_{eff,xx}(\omega, \mathbf{k} = 0)$ (blue solid curve) and $\varepsilon_{eff,yy}(\omega, \mathbf{k} = 0)$ (green dashed curve) as a function of the frequency. The host medium is air. (c) and (d) normalized amplitude of the magnetic field $H_z$ in the unit cell, when the current source is directed along $x$ and $y$ respectively, at $\omega a/c \approx 1.47$. (e) and (f) represent the real part of the electric field vector $\mathbf{E}$ in the unit cell, when the current source is directed along $x$ and $y$ respectively.

It is interesting to compare the FDFD-homogenization results with the effective parameters yielded by the well-known method of extraction based on the inversion of the reflection and transmission coefficients [16-17]. In order to obtain the required reflection and transmission coefficients (for a metamaterial with one layer thickness) we used a commercial full-wave electromagnetic simulator [53]. The computed data is also shown in Fig 8b (discrete symbols), revealing a fair agreement with our homogenization method. It may be seen that the resonance is slightly shifted to higher frequencies and that the magnetic response is weaker. The justification for these properties may be that the effective permeability extracted with our method is a parameter intrinsic to the periodic material (bulk permeability), whereas the permeability obtained from the method of extraction reported in Refs. [16-17] depends on the thickness of the metamaterial slab (in our simulation a mono-layer) as well on interface effects.

Fig. 8b shows the real part of the effective permittivity components $\varepsilon_{eff,xx}(\omega, \mathbf{k} = 0)$ and $\varepsilon_{eff,yy}(\omega, \mathbf{k} = 0)$ along the two principal directions of the structure, $x$ and $y$ respectively, as a function of frequency. It can be seen that the dielectric function along the $y$ axis ($\varepsilon_{eff,yy}$) is barely sensitive to the variation of the frequency whereas $\varepsilon_{eff,xx}$ hits a resonance at $\omega a/c \approx 1.57$. This resonance is a consequence of the coupling between the electric and magnetic fields (bianisotropy) existent in this structure. This issue will be further discussed ahead.

Figs. 8c and 8d show the normalized amplitude of the magnetic field $H_z$ (at $\omega a/c = 1.47$, and for the same material parameters as in the example of Fig 8a) when the external current source is directed along $x$ and $y$ (supposing that $\mathbf{k} = 0$), respectively. Likewise, Figs. 8e and 8f show snapshots (at $t=0$) of the electric field vector $\mathbf{E}$ in the unit cell when the current source is directed along $x$ and $y$, respectively. It is important to note that for $\mathbf{k} = 0$ the external excitation enforces a nonzero external electric field in the system, while the external magnetic induction field vanishes. Or in other words, for $\mathbf{k} = 0$ we are calculating the electric response of the metamaterial. Consistent with this property, it is seen from Fig. 8f that the electric field distribution is somehow similar to that of two electric dipoles. Fig. 8d shows that the induced current flows from one end to the other in each arm of the horseshoe, creating two magnetic fields that cancel out in all the points with coordinates $x/a = 0.5$.

On the other hand, when the current source is directed along $x$ (Figs. 8c and 8e), i.e. $\mathbf{J}_e = J_e \hat{\mathbf{u}}_x$, the magnetic field $H_z$ and the electric field are highly concentrated between the arms of the horseshoe. Even though $\mathbf{k} = 0$, such field distribution corresponds to the excitation of a magnetic resonance of the system. This occurs due to the lack of inversion symmetry of the considered material, which implies the emergence of bianisotropic effects. Thus, even though the external excitation is purely electric ($\mathbf{k} = 0$), the field distribution in Figs. 8c and 8e is mostly determined by the magnetic response of the metamaterial, since the metamaterial is operated close to the magnetic resonance (see Fig. 8a).

To demonstrate in a conclusive manner the emergence of the bianisotropic effects, we have calculated using Eq. (10) the magnetoelectric coupling parameters $\zeta_{zx}$ and $\zeta_{zy}$ of the metamaterial. The numerical simulations showed that $\zeta_{zy} \approx 0$, consistent with the symmetries of the horseshoe, specifically with the fact that the structure is invariant under the transformation $(x,y,z) \to (-x,y,z)$. On the other hand, the amplitude of the parameter $\zeta_{zx}$ may be quite significant.

Fig. 9 represents $\text{Im}\{\zeta_{zx}\}$ as a function of frequency (for a lossless metamaterial the tensor $\bar{\bar{\zeta}}$ is purely imaginary [31]), showing that it exhibits the same behavior as the magnetic permeability, hitting a resonance at the same frequency ($\omega a/c = 1.47$). We have also plotted in Fig. 9 the frequency dependence of $\varepsilon_{eff,xx}(\omega, \mathbf{k}=0)$, which, from Eq. (8), is related to the local parameters through the relation $\varepsilon_{eff,xx}(\omega, \mathbf{k}=0) = \varepsilon_{xx} + \frac{1}{c^2}\zeta_{zx}^2 \frac{1}{\mu_{eff}}$, where $\varepsilon_{xx}$ is the relative *local* permittivity along *x*. Interestingly, the previous formula predicts that, in presence of bianisotropic effects, $\varepsilon_{eff,xx}$ has a resonance at the frequency where the permeability $\mu_{eff}$ vanishes. This is confirmed by Fig. 9, which shows that at $\omega a/c \approx 1.55$ the magnetic permeability vanishes and the *xx* component of the nonlocal dielectric function has a pole.

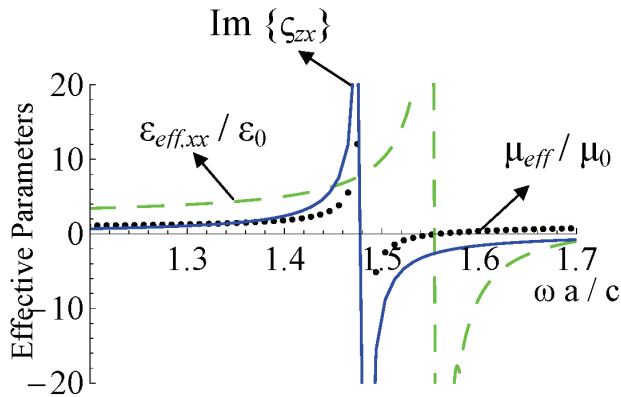

Fig. 9 (Color online) Nonlocal dielectric function along the *x*-direction (green dashed line), effective permeability (black dotted line) and magnetoelectric parameter (blue solid line) for the horseshoe geometry of Fig. 8a.

It is important to stress that the extraction of the material parameters ($\zeta_{zx}$, $\mu_{eff}$, $\varepsilon_{xx}$, and $\varepsilon_{yy}$) is based on the assumption that the effective medium can be accurately described by the bianisotropic constitutive relations [Eq. (6)]. In these conditions, the nonlocal dielectric function is necessarily a quadratic form of the wave vector, as follows from Eq. (5). In particular, it should be clear that the magnetic permeability $\mu_{eff}$ should completely determine the spatial dispersion effects of second order, or equivalently it should univocally determine the second order derivatives of the nonlocal dielectric function $\overline{\overline{\varepsilon}}_{eff}$ with respect to $\mathbf{k}$. It may be easily verified from Eq. (5) (by calculating the derivatives $\partial^2/\partial k_x^2$ and $\partial^2/\partial k_x \partial k_y$ at the origin) that $\mu_{eff}$ should verify, besides Eq. (7), the following formulas:

$$\frac{\mu_{eff}^{(2)}}{\mu_0} = \frac{1}{1 - \left(\frac{\omega}{c}\right)^2 \frac{1}{2\varepsilon_0} \left.\frac{\partial^2 \varepsilon_{eff,xx}}{\partial k_y^2}\right|_{\mathbf{k}=0}} \qquad (16a)$$

$$\frac{\mu_{eff}^{(3)}}{\mu_0} = \frac{1}{1 + \left(\frac{\omega}{c}\right)^2 \frac{1}{\varepsilon_0} \left.\frac{\partial^2 \varepsilon_{eff,xy}}{\partial k_x \partial k_y}\right|_{\mathbf{k}=0}} \qquad (16b)$$

The above formulas should be regarded as consistency conditions of the bianisotropic model, i.e. if the material is in fact local it must verify $\mu_{eff}^{(1)} = \mu_{eff}^{(2)} = \mu_{eff}^{(3)}$, where $\mu_{eff}^{(1)}$ is the magnetic permeability extracted using Eq. (7).

In order to check this, we numerically calculated the magnetic permeability (for the configuration of Fig. 8a) using the extraction formulas $\mu_{eff}^{(1)}$, $\mu_{eff}^{(2)}$ and $\mu_{eff}^{(3)}$. The result is plotted in Fig. 10 where it can be seen that the consistency relations are not satisfied by the horseshoe configuration, since the curves describing each of the extraction formulas do not agree. Moreover, the curves associated with $\mu_{eff}^{(2)}$ and $\mu_{eff}^{(3)}$ exhibit a resonance with a nonphysical dispersion at $\omega a/c = 1.6$. This unsettling result implies that the magnetic permeability of the structure depends on the direction of propagation, and thus the response of the horseshoe particle cannot be fully described by the assumed bianisotropic relations [Eq. (6)]. Indeed, it must be emphasized that a truly local material (with $\zeta_{zx}, \mu_{eff}, \varepsilon_{xx}$, and $\varepsilon_{yy}$ independent of **k**) should verify $\mu_{eff}^{(1)} = \mu_{eff}^{(2)} = \mu_{eff}^{(3)}$. Therefore, in general the considered metamaterial should be regarded as nonlocal (in the sense that it cannot be described by the local constitutive relations (6)).

The origin of the spatially dispersive response of the metamaterial may be related to the excitation of the electric quadrupole moment, which is known to have strength comparable to that of the magnetic dipole, in other metamaterials with a topology similar to that of the horseshoe [54, 55]. The excitation of the electric quadrupole is a consequence of the fact that the induced electric current does not form a closed loop [54] (Figs. 8c and 8d). Indeed, we numerically verified that to a very good approximation $\mu_{eff}^{(1)} = \mu_{eff}^{(2)} \approx \mu_{eff}^{(3)}$ in the examples considered in sections III.A and III.C (inclusions with circular cross-section; the circular geometry forces the induced current to circulate in a loop).

Another reason for the lack of consistency between $\mu_{eff}^{(1)}$, $\mu_{eff}^{(2)}$ and $\mu_{eff}^{(3)}$ in the geometry of Fig. 8a may be the fact that the volume density of the horseshoe inclusions in that example is too high to expect a local response of the material (these high densities of particles are typical of metamaterials, and are necessary to have a strong magnetic response). When working with densely packed crystals the expansion of the dielectric function must include additional terms which are second-order with respect to **k** and possibly other terms of even higher orders. Physically, those terms are related to the higher-order multipole moments that are not taken into account by the bianisotropic constitutive relations [Eq. (6)].

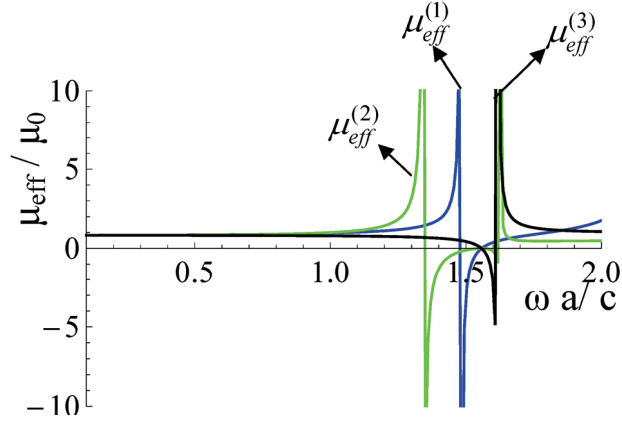

Fig. 10 (Color online) Real part of the effective permittivity $\mu_{eff}$ as a function of the normalized frequency $\omega a/c$ considering different formulas for the extraction of $\mu_{eff}$.

We have also studied the influence of the permittivity of the horseshoe inclusions on the magnetic response of the particles. Fig. 11 depicts the required permittivity for the particles in order that the magnetic resonance occurs at a certain specified frequency $\omega_r a/c$ (in these plots the permittivity of the nanoparticles is regarded as a free-parameter; the dimensions of the particle are kept invariant; the star-shaped symbols were obtained using the homogenization method proposed here [using Eq. (7)], whereas the circle-shaped symbols were obtained using the retrieval process of Refs. [16-17]). The results show that to obtain very subwavelength resonant particles ($\omega_r a/c << 1$) the absolute value of the real part of the permittivity of the inclusions $-\varepsilon'_r$ must be relatively small. Specifically, in order to obtain ultrasubwavelength particles it is necessary that the skin depth of the metal, $\delta$, be comparable or larger than the thickness of the nanoantennas. This is clearly seen in the inset of Fig. 11, which represents the normalized resonant wavelength $\lambda_r/a$ as a function of the normalized skin depth $\delta/b$ of the horseshoe configuration.

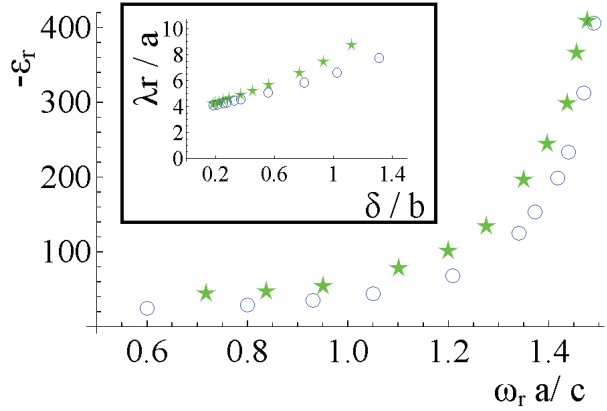

Fig. 11 (Color online) Required permittivity $\varepsilon_r$ for the horseshoe inclusion as a function of the normalized resonant frequency $\omega_r a / c$. The inset shows the normalized resonant wavelength $\lambda_r / a$ as a function of the normalized skin depth $\delta / b$ of the metal. The circle-shaped symbols correspond to the values extracted with the method of inversion of the reflection and transmission coefficients [16-17] and the star-shaped symbols were extracted with the FDFD method.

## IV. CONCLUSION

In this work, we developed a FDFD implementation of the nonlocal homogenization approach proposed in [20]. We have demonstrated that the proposed numerical method may be an excellent solution to solve the homogenization problem, yielding very accurate results, and allowing for the computation of the effective parameters of metamaterials formed by dielectric and metallic inclusions with arbitrary shapes, taking into account both the effect of loss and frequency dispersion. We compared our homogenization approach with other homogenization methods, demonstrating not only its accuracy but also its generality. In particular, we have studied with details the electrodynamics of arrays of horseshoe shaped nanoparticles, and emphasized the fact that such metamaterials may be characterized by significant magneto-electric coupling, as well as nonlocal effects that cannot be described by the usual bianisotropic constitutive relations. The developed formalism can be easily extended to fully three-dimensional structures.

## ACKNOWLEDGEMENT

This work was funded by Fundação para Ciência e a Tecnologia under project PDTC/EEA-TEL/71819/2006. J. C. acknowledges financial support by Fundação para a Ciência e a Tecnologia under the fellowship SFRH/BD/36976/2007. The authors gratefully thank Dr. H. Wallén, and Prof. A. Sihvola for kindly providing the numerical data of their work [48].